\documentclass[a4paper,11pt]{article}
\usepackage{pos}

\newcommand{\eps}{\epsilon}
\newcommand{\pvec}{\mathbf{p}}
\newcommand{\qvec}{\mathbf{q}}
\newcommand{\xvec}{\mathbf{x}}
\newcommand{\mPsi}{m_{\Psi}}
\newcommand{\Qt}{\widetilde{Q}^{\,2}}
\newcommand{\wt}{\widetilde{\omega}}
\newcommand{\as}{\alpha_s}
\newcommand{\CF}{C_F}
\newcommand{\xiavg}[1]{\left\langle \xi^{#1}\right\rangle}

\title{Parton physics from a heavy-quark operator product expansion:\\
Dynamical lattice QCD calculation of moments of the pion and kaon light-cone distribution amplitudes}
\ShortTitle{HOPE for pion/kaon LCDA moments}

\author*[a]{S.-P. Alex Chang}
\author[b]{William Detmold}
\author[c]{Anthony V. Grebe}
\author[b]{Mat\'ias Guti\'errez-Escobari}
\author[d]{Issaku Kanamori}
\author[a,e]{C.-J. David Lin}
\author[b]{Robert J. Perry}
\author[f]{Yong Zhao}

\affiliation[a]{Institute of Physics, National Yang Ming Chiao Tung University,\\
Hsinchu 30010, Taiwan}
\affiliation[b]{Center for Theoretical Physics - A Leinweber Institute, Massachusetts Institute of Technology,\\
Cambridge, MA 02139, USA}
\affiliation[c]{Department of Physics and Maryland Center for Fundamental Physics, University of Maryland,\\
College Park, MD 20742, USA}
\affiliation[d]{RIKEN Center for Computational Science,\\
Kobe 650-0047, Japan}
\affiliation[e]{Centre for High Energy Physics, Chung-Yuan Christian University, \\
Taoyuan 320314, Taiwan}
\affiliation[f]{Physics Division, Argonne National Laboratory,\\
Lemont, IL 60439, USA}

\emailAdd{
s44930e0@gmail.com, 
wdetmold@mit.edu,
matiasge@mit.edu,
agrebe@mit.edu,
kanamori-i@riken.jp,
dlin@nycu.edu.tw, 
perryrobertjames@gmail.com, 
yong.zhao@anl.gov
}

\abstract{
The light-cone distribution amplitude (LCDA) is a fundamental non-perturbative quantity for understanding hadron structure and exclusive scattering processes.
We report on our calculation of the pion and kaon LCDAs using the heavy-quark operator product expansion (HOPE) framework. This method employs an OPE analysis of hadronic amplitudes through the inclusion of a fictitious valence heavy quark.
In these proceedings, we report progress on the determination of the first three nontrivial Mellin moments of the kaon LCDAs from dynamical lattice QCD calculations, and we summarize the recently published continuum-limit result for the pion fourth Mellin moment obtained in the quenched approximation, thereby demonstrating the feasibility of the HOPE method for accessing higher moments.
}

\FullConference{The 42nd International Symposium on Lattice Field Theory (LATTICE2025)\\
2--8 November 2025\\
Tata Institute of Fundamental Research, Mumbai, India}

\usepackage{mfirstuc} 
\usepackage{todonotes} 
\newcommand{\addReviewer}[2]{
  \expandafter\newcommand\csname #1\endcsname[1]{{\textbf{ \color{#2} \capitalisewords{#1}:\,##1}}}
  \expandafter\newcommand\csname #1cor\endcsname[2]{{\color{#2} \capitalisewords{#1}:\,\st{##1}{\textbf{##2}}}}
  \expandafter\newcommand\csname #1color\endcsname{#2}
  \expandafter\newcommand\csname #1todo\endcsname[1]{{\todo[inline,color=white!70!#2, caption={}]{\textbf{\capitalisewords{#1}}: ##1}}}
}

\usepackage{tikz, marginnote} 
\newcommand{\checkedby}[1]{
\ifdefined\CROSSCHECKS
  \marginnote{
    \begin{tikzpicture}
      \foreach \x [count=\xi] in {#1} {
         \node[shape=circle,inner sep=0mm,
         minimum size=2mm,
         fill=\csname \x color\endcsname] at (\xi*3mm,0) {};
       }
    \end{tikzpicture}
  }
\else
\fi
}

\usepackage{soul,color}

\definecolor{chromeyellow}{rgb}{1.0, 0.65, 0.0}
\definecolor{DodgeBlue}{rgb}{0.118, 0.565,1.000}
\definecolor{asparagus}{rgb}{0.53, 0.66, 0.42}
\definecolor{cadmiumgreen}{rgb}{0.0, 0.42, 0.24}
\definecolor{blue(ryb)}{rgb}{0.01, 0.28, 1.0}
\definecolor{periwinkle}{RGB}{181, 146, 203}
\definecolor{turquoiseblue}{rgb}{0.02, 0.55, 0.55}
\definecolor{green1}{RGB}{50,205,50}
\definecolor{amethyst}{rgb}{0.6, 0.4, 0.8}
\definecolor{indianred}{RGB}{205,92,92}
\definecolor{applegreen}{rgb}{0.55,0.71,0.0}
\definecolor{teal}{RGB}{66, 245, 197}
\definecolor{olivegreen}{HTML}{3C8031}

\addReviewer{willd}{orange}
\addReviewer{david}{olivegreen}
\addReviewer{anthony}{blue}
\addReviewer{alex}{cadmiumgreen}
\addReviewer{issaku}{magenta}
\addReviewer{yong}{cyan}
\addReviewer{robert}{turquoiseblue}

\newcommand{\str}[1]{}  

\begin{document}
\maketitle

\section{Introduction}

QCD factorization theorems ensure the separation of short-distance kernels, computable in perturbation theory, and universal hadronic matrix elements that encode long-distance dynamics.  For a wide class of high-energy exclusive processes, the relevant long-distance input is the leading-twist light-cone distribution amplitude (LCDA) of the participating meson
\cite{Efremov:1980qk,Lepage:1980fj,Braun:2003rp,Braun:2007wv}.

The LCDA is defined through the matrix element of a gauge-invariant operator involving a light-like separation,
\begin{equation}
\langle 0 | \bar q_f(z)\,\gamma^\mu\gamma_5\,W[z,-z]\,q_{f^\prime}(-z) | M(p)\rangle
= i f_M p^\mu \int_{-1}^{1} d\xi\, e^{i\xi\, p\!\cdot\! z}\,\phi_M(\xi,\mu),
\label{eq:lcda_def}
\end{equation}
where $p^\mu$ is the four-momentum of the pseudoscalar meson $M$ and $f_M$ is its decay constant. For the pion, the flavor quantum numbers of the quark fields are $f = u$, $f^\prime = d$, while for the kaon the flavor quantum numbers are $f = s$, $f^\prime = d$. The Wilson line
$W[z,-z]$ connects $-z$ and $z$ along a light-like path, and $\mu$ denotes the renormalization scale.  In
light-cone gauge, $\phi_M(\xi,\mu)$ admits a partonic interpretation as the amplitude for the meson to be in a valence
quark--antiquark state with momentum fractions $(1\pm\xi)/2$.

At leading twist, conformal symmetry motivates an expansion of the meson LCDA in Gegenbauer polynomials $C_n^{(3/2)}$ \cite{Braun:2003rp,Braun:2007wv},
\begin{equation}
\phi_M(\xi,\mu)=\frac{3}{4}\left(1-\xi^2\right)\sum_{n=0}^{\infty}\phi_{n,M}(\mu)\,C_n^{(3/2)}(\xi),
\label{eq:gegenbauer_expansion}
\end{equation}
where the Gegenbauer moments $\phi_{n,M}(\mu)$ evolve multiplicatively with $\mu$ according to their known anomalous dimensions at one loop.
As $\mu\to\infty$, higher-$n$ moments are suppressed because of the properties of these one-loop anomalous dimensions, and  the LCDA approaches the asymptotic form~\cite{Lepage:1980fj},
\begin{equation}
\lim_{\mu\to\infty}\phi_M(\xi,\mu)=\frac{3}{4}\left(1-\xi^2\right).
\label{eq:asymptotic_lcda}
\end{equation}
At experimentally accessible momentum scales, exclusive observables like meson--photon transition form factors \cite{Gronberg:1997fj,Aubert:2009mc,Uehara:2012ag} remain sensitive to non-asymptotic components of the LCDA, motivating non-perturbative constraints at hadronic scales.

A central challenge is that Eq.~\eqref{eq:lcda_def} involves a light-like separation and cannot be accessed directly in Euclidean lattice QCD.
This has motivated indirect strategies to extract light-cone information from the lattice.
Large-momentum effective theory (LaMET) calculates the $x$-dependence within a moderate range from large-momentum quasi-distributions defined from Euclidean matrix elements~\cite{Ji:2013dva,Ji:2014gla,Ji:2020ect,Hua:2022kcm}, while approaches based on short-distance expansions~\cite{Braun:2007wv,Ma:2017pxb,Radyushkin:2017cyf,Bali:2019dqc} aim to reconstruct the $x$-dependence within the entire range of $[0,1]$ from similar matrix elements through phenomenological fitting; in practice, finite hadron momenta and power corrections make the endpoint regions particularly challenging for all approaches, while in fitting one also needs to control and quantify the model uncertainties. Therefore, complementary observables that do not rely on resolving the full $\xi$-dependence remain valuable~\cite{Cichy:2018mum}.

Global constraints on the LCDA are given by the Mellin moments,
\begin{equation}
\xiavg{n}_M(\mu)=\int_{-1}^{1} d\xi\,\xi^n\,\phi_M(\xi,\mu),
\label{eq:mellin_def}
\end{equation}
which summarize the shape information of the LCDA and can be computed using lattice QCD.
Mellin moments can be related to matrix elements of local twist-two operators, but beyond the lowest moments the conventional method of performing lattice computations of matrix elements of local operators becomes increasingly difficult due to power-divergent mixing and operator proliferation
\cite{Martinelli:1987un,DelDebbio:1999zq,Detmold:2005gg,Arthur:2010xf}. Recently, additional ideas to address these difficulties for parton observables have been explored, including current--current and Wilson-line correlator approaches based on short-distance OPEs \cite{Braun:2007wv,Ma:2017pxb,Izubuchi:2018srq,Gao:2022vyh},
as well as lattice implementations using the Feynman--Hellmann theorem \cite{Can:2020zku} and gradient-flow constructions \cite{Shindler:2023gsa}.

In these proceedings we focus on a moment-oriented strategy, the heavy-quark operator product expansion (HOPE).
In the HOPE strategy, one introduces a fictitious valence heavy quark to provide a hard scale and studies a Euclidean current--current correlator that admits a short-distance expansion with continuum-like operator mixing \cite{Detmold:2005gg,Detmold:2021uru}.
The matching coefficients are known at one loop \cite{Detmold:2021uru}, enabling controlled fits for Mellin moments.  
Building on recent HOPE developments and applications \cite{Detmold:2021uru,Detmold:2021qln,Perry:2023hope,Perry:2025pos,Detmold:2025lyb}, we present the status of the first three nontrivial moments of the dynamical ($N_f=2+1$) kaon and briefly summarize a recently published determination of the pion fourth Mellin moment obtained in the quenched approximation\cite{Detmold:2025lyb}.
\section{HOPE method}

We introduce a heavy valence quark field $\Psi$ with mass $\mPsi$ and define the heavy--light axial currents \cite{Detmold:2021uru}
\begin{equation}
J^{f}_\mu(x)\equiv \bar\Psi(x)\,\gamma_\mu\gamma_5\,q_f(x)
+ \bar q_f(x)\,\gamma_\mu\gamma_5\,\Psi(x),
\label{eq:axial_currents}
\end{equation}
where $q_f$ denotes the light or strange quark field.
The Minkowski hadronic tensor is
\begin{equation}
V_{\mu\nu}^{M}(q,p)= i\int d^4x\,e^{iq\cdot x}\,
\langle 0|\,T\{J^{f}_\mu(x)\,J^{f'}_\nu(0)\}\,|M(p)\rangle .
\label{eq:V_mink}
\end{equation}
and lattice computations access the corresponding Euclidean correlator in the regime of space-like separation.
The short-distance OPE is expressed in terms of the kinematic variables,
\begin{equation}
Q^2\equiv -q^2,\qquad
\Qt \equiv Q^2+\mPsi^2,\qquad
\wt \equiv \frac{2\,p\!\cdot\! q}{\Qt},
\label{eq:kin}
\end{equation}
The hadronic tensor can be parametrized in terms of an invariant amplitude as
\begin{equation}
V_{\mu\nu}^{M}(q,p)
= -\frac{2 i f_M\,\eps_{\mu\nu\rho\sigma}q^\rho p^\sigma}{\Qt}\;
\mathcal{V}_M(\wt,\Qt,\mu)
+ \text{higher twist}.
\label{eq:Vproj}
\end{equation}
At one-loop order,  the HOPE Wilson coefficients are known, and it is convenient to express the expansion in terms of Gegenbauer moments $\phi_{n,M}(\mu)$ \cite{Detmold:2021uru,Detmold:2021qln}:
\begin{equation}
\mathcal{V}_M(\wt,\Qt,\mu)
= \sum_{n=0}^{\infty} F_n(\wt,\Qt,\mu,\mPsi)\,\phi_{n,M}(\mu),
\qquad
F_n = F_n^{(0)} + \frac{\as(\mu)\CF}{4\pi}F_n^{(1)}+\cdots .
\label{eq:one_loop}
\end{equation}
Here $F_n$ are Wilson coefficients that can be computed in QCD perturbation theory, with $F_{n}^{(0,1)}$ being the tree-level and one-loop results.
The Mellin and Gegenbauer moments are related by comparing the tree-level series in Eq.~\eqref{eq:gegenbauer_expansion} with Eq.~\eqref{eq:mellin_def}~\cite{Detmold:2021qln,Detmold:2025lyb}.\\

\section{Lattice calculation: correlators and analysis strategy}
We follow the HOPE strategy of Refs.~\cite{Detmold:2005gg,Detmold:2021uru}. 
In this work, calculation with ($N_f=2+1$) dynamical configurations is performed for the kaon on the CLS ensembles listed in Table~\ref{tab:ensembles}.

\begin{table}
\centering
\caption{Lattice details for dynamical ensembles, generated by the CLS collaboration, employed in calculation of the first three nontrivial Mellin moments of the kaon. Further details about these ensembles can be found in Ref.~\cite{Bali:2022cls}.}
\label{tab:ensembles}
\begin{tabular}{l c c c c c c c}
\hline
Name & $(L/a)^3 \times T/a$ & $\beta$ & $a$ (fm) & $\kappa_{\rm light}$ & $\kappa_{\rm strange}$ & $m_\pi$ (MeV) & $m_K$ (MeV) \\
\hline
B451 & $32^3 \times 64$  & 3.46 & 0.075 & 0.136981 & 0.136409 & 422 & 577 \\
N305 & $48^3 \times 128$ & 3.7  & 0.049 & 0.137025 & 0.136676 & 428 & 584 \\
B452 & $32^3 \times 64$  & 3.46 & 0.075 & 0.137046 & 0.136378 & 352 & 548 \\
N304 & $48^3 \times 128$ & 3.7  & 0.049 & 0.137079 & 0.136665 & 353 & 558 \\
N450 & $48^3 \times 128$ & 3.46 & 0.075 & 0.137099 & 0.136353 & 287 & 528 \\
\hline
\end{tabular}
\end{table}

We work in a mixed time--momentum representation and define
\begin{equation}
R_{\mu\nu}^M(\tau;\pvec,\qvec)
\equiv
\int d^3\mathbf{z}\,e^{i\qvec\cdot\mathbf{z}}\,
\langle 0|\,T\!\left\{J^f_\mu(\tau,\mathbf{z})\,J^{f'}_\nu(0,\mathbf{0})\right\}\,|M(\pvec)\rangle ,
\label{eq:R_def_matrixelement}
\end{equation}
which is the matrix element of the heavy--light axial current product at Euclidean separation $\tau$.
On the lattice, $R_{\mu\nu}^M(\tau;\pvec,\qvec)$ is obtained from ratios of three- and two-point functions after removing excited-state contamination, as described below.

\subsection{Two- and three-point functions and the ratio construction}
We determine the ground-state energy $E_0(\pvec)$ and overlap factor
$Z_{0,M}(\pvec)\equiv \langle 0|O_M(0,\mathbf{0})|M(\pvec)\rangle$
from the meson two-point function
\begin{equation}
C_M(\tau,\pvec)=\sum_{\xvec} e^{i\pvec\cdot\xvec}\,
\langle 0|\,O_M(\tau,\xvec)\,O_M^\dagger(0,\mathbf{0})\,|0\rangle
=\sum_{n}\frac{|Z_{n,M}(\pvec)|^2}{2E_n(\pvec)}\,e^{-E_n(\pvec)\tau},
\label{eq:C2_spec_31}
\end{equation}
We then compute the HOPE three-point correlator with two heavy--light currents,
\begin{equation}
C^{\mu\nu}_M(\tau_e,\tau_m;\pvec_e,\pvec_m)=
\sum_{\xvec_e,\xvec_m} e^{i\pvec_e\cdot\xvec_e}\,e^{i\pvec_m\cdot\xvec_m}\,
\langle 0|\,T\{J_A^\mu(\tau_e,\xvec_e)\,J_A^{\nu}(\tau_m,\xvec_m)\,O_M^\dagger(0,\mathbf{0})\}\,|0\rangle ,
\label{eq:C3_def_31}
\end{equation}
and introduce the definitions,
\begin{equation}
\tau \equiv \tau_e-\tau_m,\qquad
\pvec \equiv \pvec_e+\pvec_m,\qquad \qvec \equiv \frac{\pvec_e-\pvec_m}{2}.
\label{eq:kin_31}
\end{equation}
We then construct the ratio.
\begin{equation}
\mathcal{R}^{\mu\nu}_M(\tau_e,\tau_m;\pvec,\qvec)\equiv
2E_0(\pvec)\frac{C^{\mu\nu}_M(\tau_e,\tau_m;\pvec_e,\pvec_m)}
{Z_{0,M}\,e^{-E_0(\pvec)\,(\tau_e+\tau_m)/2}}.
\label{eq:R_def_31}
\end{equation}
In the limit of $\tau_e\to\infty$ for fixed $\tau$, this ratio approaches the hadronic matrix element of interest, given by equation~\eqref{eq:R_def_matrixelement}.

\begin{figure}[htb]
  \centering
  \includegraphics[width=0.65\textwidth]{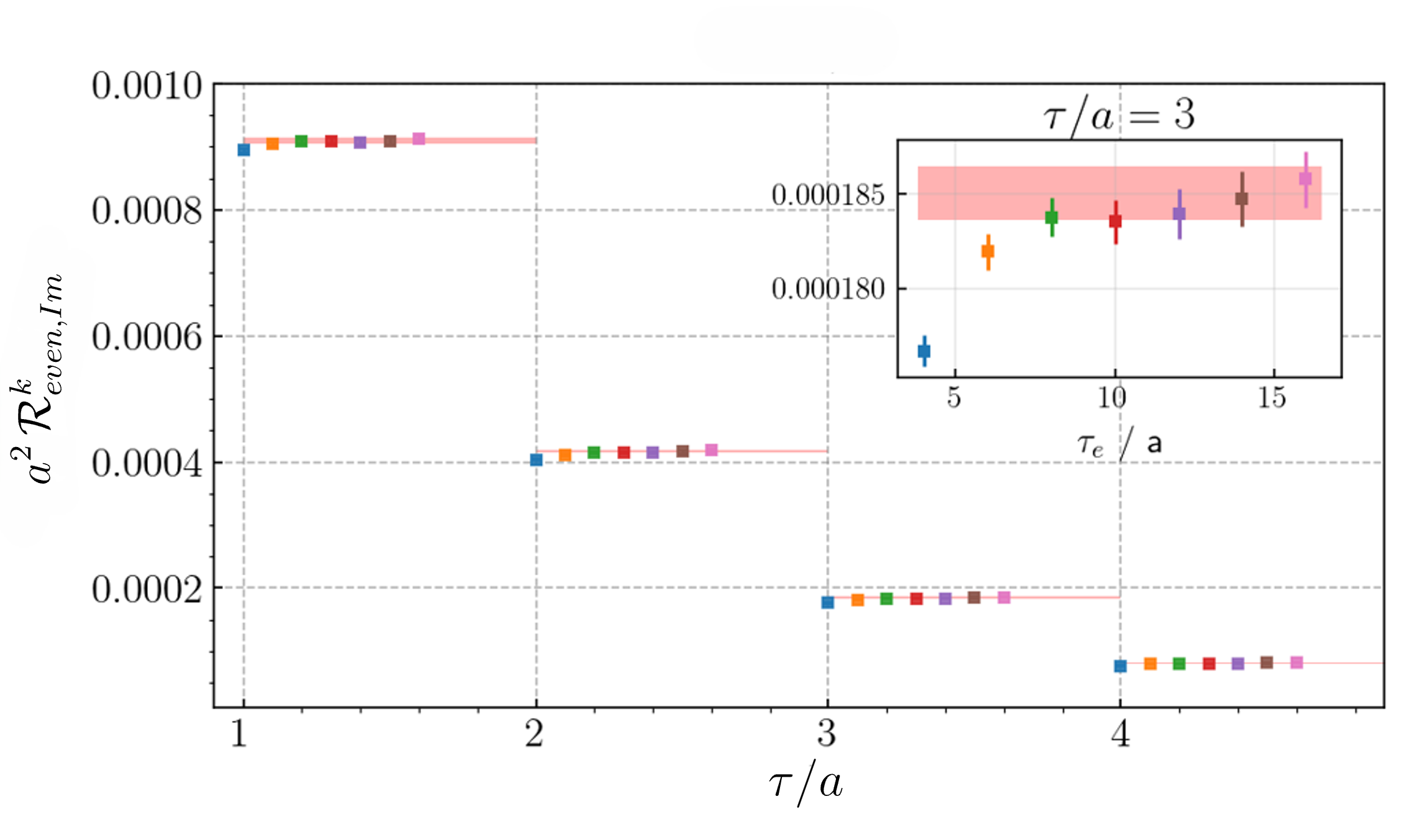} 
  \caption{Excited-state control for the kaon Euclidean-time ratio in the even channel. Shown are $a^{2}R_{\mathrm{even,Im}}$ as a function of the current separation $\tau/a$ with $\tau\equiv \tau_e-\tau_m$. For each $\tau/a$, the colored points correspond to different choices of the Euclidean time extent $\tau_e/a$ (equivalently $\tau_+$) used in the ratio construction, illustrating the approach to a plateau as $\tau_e$ increases.
The inset highlights the $\tau/a=3$ case versus $\tau_e/a$, and the shaded band indicates the asymptotic value obtained from an exponential extrapolation to $\tau_e\to\infty$ (see Eq.~\eqref{eq:R_excited_model}).}

  \label{fig:excited}
\end{figure}

\subsection{Excited-state contamination and extrapolation}
At finite Euclidean time extent, $\mathcal{R}^{\mu\nu}_M(\tau_e,\tau_m;\pvec,\qvec)$ receives exponentially suppressed excited-state contributions.
We vary $\tau_e$ at fixed $\tau=\tau_e-\tau_m$ and remove excited-state contamination by extrapolating to the large-$\tau_e$ limit.
We model the leading correction as
\begin{equation}
\mathcal{R}^{\mu\nu}_M(\tau_e,\tau_m;\pvec,\qvec)
=
R^{\mu\nu}_M(\tau;\pvec,\qvec)\,
\Big[1 + a^{\mu\nu}(\tau;\pvec,\qvec)\,e^{-\Delta E\,\tau_e}
+ \mathcal{O}\!\left(e^{-\Delta E'\,\tau_e}\right)\Big],
\label{eq:R_excited_model}
\end{equation}
where $\Delta E\equiv E_1(\pvec)-E_0(\pvec)$ and $\Delta E'>\Delta E$ denotes the gap to higher excited states.
The extrapolated $R^{\mu\nu}_M(\tau;\pvec,\qvec)$ is taken as the excited-state--removed input to the HOPE fits.
An example of a projected observable derived from $R^{\mu\nu}_M$, defined later in Eq.~\eqref{eq:R_split_def}, is shown in Fig.~\ref{fig:excited}.

\subsection{Numerical details}

We focus on the antisymmetric component in the Lorentz indices,
\begin{equation}
V^{M}(q,p)\equiv \frac{1}{2}\left[V_{12}^{M}(q,p)-V_{21}^{M}(q,p)\right],
\label{eq:Vproj_transverse}
\end{equation}
and choose kinematics with $p_3=0$.  For this choice, the prefactor
$i\,\eps_{\mu\nu\rho\sigma}q^\rho p^\sigma$ in Eq.~\eqref{eq:Vproj} is purely imaginary.
As a result, the zeroth-moment contribution enters only through the imaginary part, while higher moments can induce a nonzero real part through $p\!\cdot\! q$. Further details on this point can be found in Ref.~\cite{Detmold:2021qln}.

In the isospin-symmetric limit, the pion LCDA is symmetric and all odd moments vanish, $\xiavg{2k+1}_\pi=0$.
For the kaon, odd moments are generally nonzero, and we separate even- and odd-moment contributions by combining data at $q$ and $-q$ \cite{Perry:2025pos}:
\begin{equation}
V^{M}_{\mathrm{even/odd}}(p,q)=\frac{1}{2}\Big[V^{M}(p,q)\pm V^{M}(p,-q)\Big].
\label{eq:evenodd}
\end{equation}

To compare directly with the lattice observable in the time--momentum representation, we transform the hadronic tensor $V^M$ to time--momentum space via an inverse Fourier transform.
\begin{equation}
R^{M}_{\mathrm{even/odd}}(\tau;\pvec,\qvec)=\frac{1}{2\pi}\int_{-\infty}^{\infty} dq_4\;
e^{-iq_4\tau}\,V^{M}_{\mathrm{even/odd}}(p,q).
\label{eq:R_from_V_invFT}
\end{equation}
We define two contributions according to whether the inverse transform is applied to the real or imaginary part of $V^M$,
\begin{align}
R^{M}_{\mathrm{even/odd},\,\mathrm{Re}}(\tau;\pvec,\qvec)
&\equiv \frac{1}{2\pi}\int_{-\infty}^{\infty} dq_4\;
e^{-iq_4\tau}\,\mathrm{Re}\!\left[V^{M}_{\mathrm{even/odd}}(p,q)\right],\nonumber\\
R^{M}_{\mathrm{even/odd},\,\mathrm{Im}}(\tau;\pvec,\qvec)
&\equiv \frac{1}{2\pi}\int_{-\infty}^{\infty} dq_4\;
e^{-iq_4\tau}\,\mathrm{Im}\!\left[V^{M}_{\mathrm{even/odd}}(p,q)\right].
\label{eq:R_split_def}
\end{align}
Altogether we construct four channels, $R^{M}_{\mathrm{even/odd},\,\mathrm{Re/Im}}$.
All channels depend on the decay constant and the heavy-quark mass through the overall kinematic prefactors and Wilson coefficients.
However, for our choice of kinematics, the even--Im channel is dominated by the decay constant at leading order, while the even--Re and odd channels suppress this leading contribution and are therefore more sensitive to higher Mellin moments.
In particular, the even--Re channel provides sensitivity to the second moment, and the odd channels provide sensitivity to the first and third moments.

\begin{figure}[h!]
  \centering
  \includegraphics[width=0.9\textwidth]{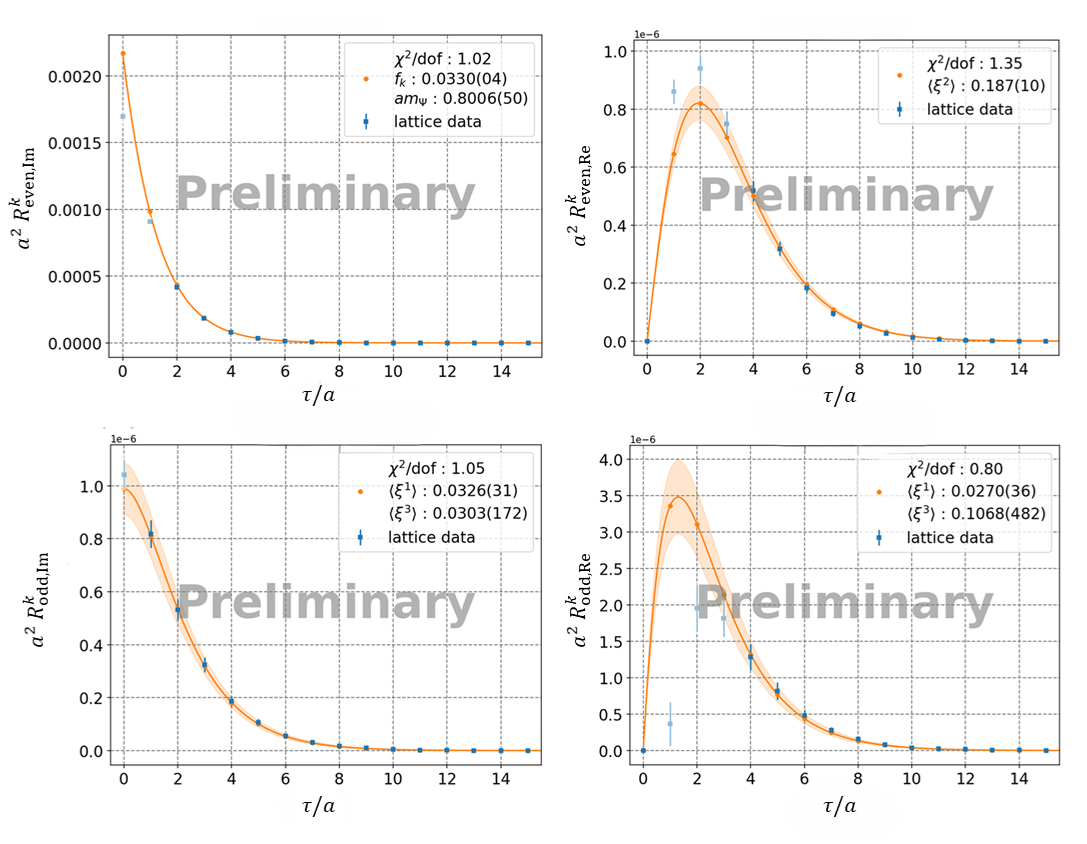}
  \caption{Kaon one-loop HOPE fits to the excited-state--removed ratios, shown in the even/odd sectors and for the real and imaginary channels.}
  \label{fig:HadronicAmplitudes}
\end{figure}

\section{Preliminary results}

\begin{figure}[h!]
  \centering
  \begin{minipage}[t]{0.49\textwidth}
    \centering
    \includegraphics[width=\linewidth]{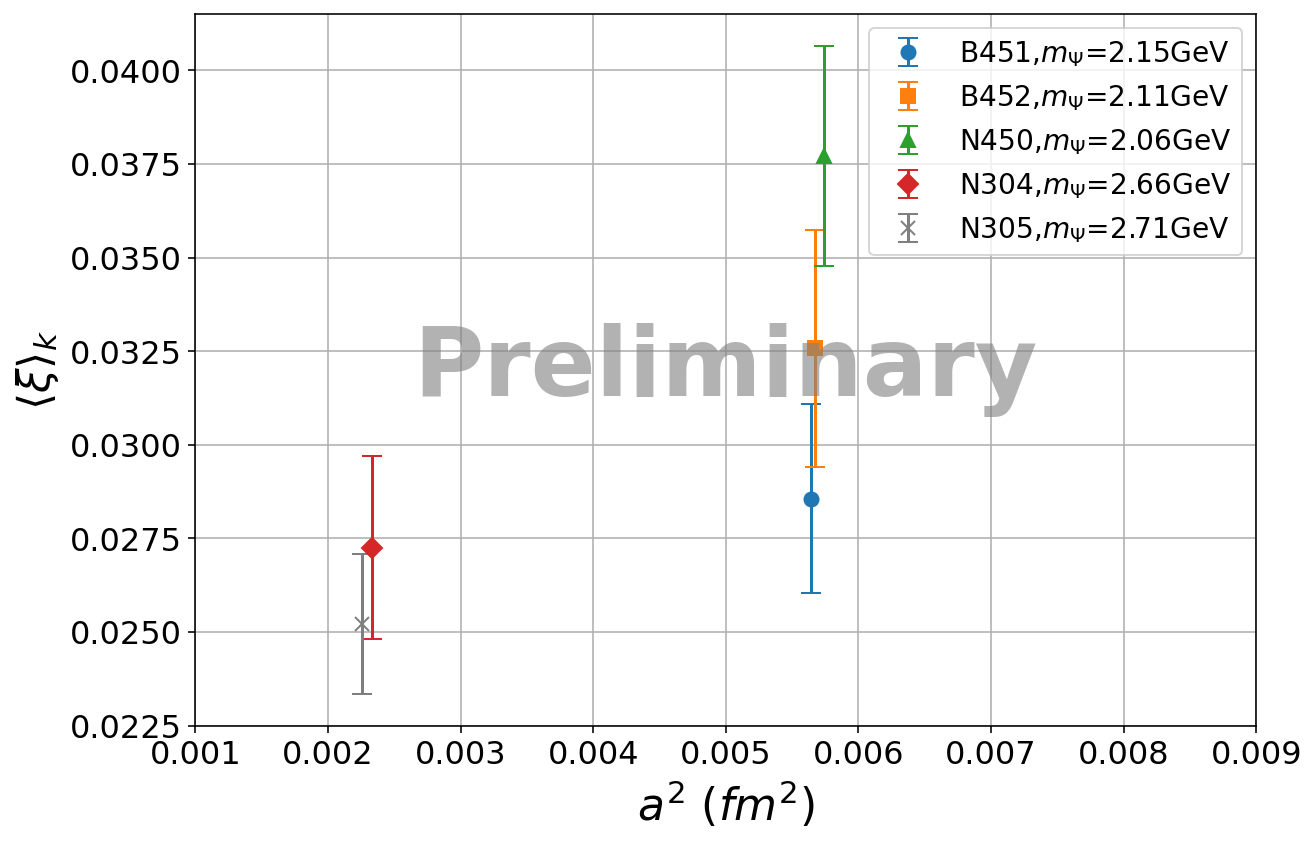}
  \end{minipage}\hfill
  \begin{minipage}[t]{0.49\textwidth}
    \centering
    \includegraphics[width=\linewidth]{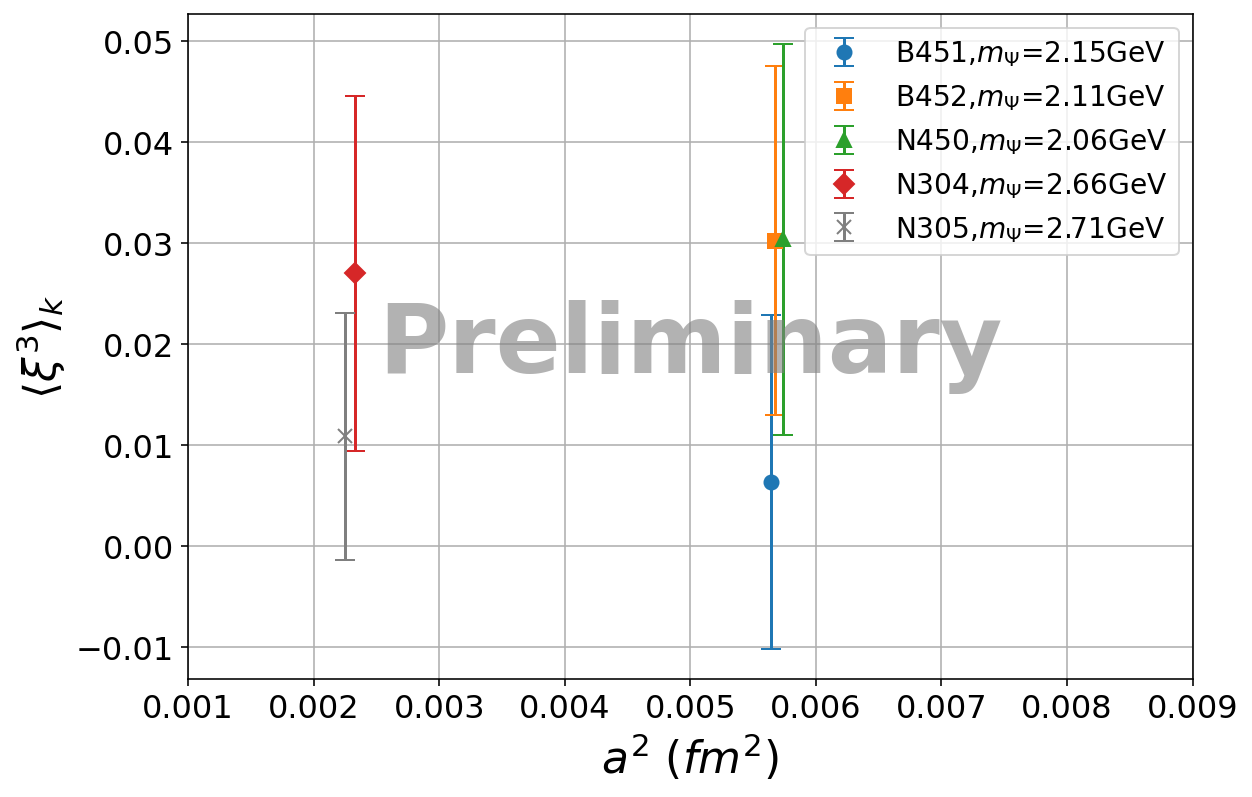}
  \end{minipage}
  \caption{Preliminary kaon results for the first (left) and third (right) Mellin moments shown versus $a^2$ for different ensembles.and heavy-quark masses.}
  \label{fig:mom13_a2}
\end{figure}

\begin{figure}[h!]
  \centering
  \begin{minipage}[t]{0.49\textwidth}
    \centering
    \includegraphics[width=\linewidth]{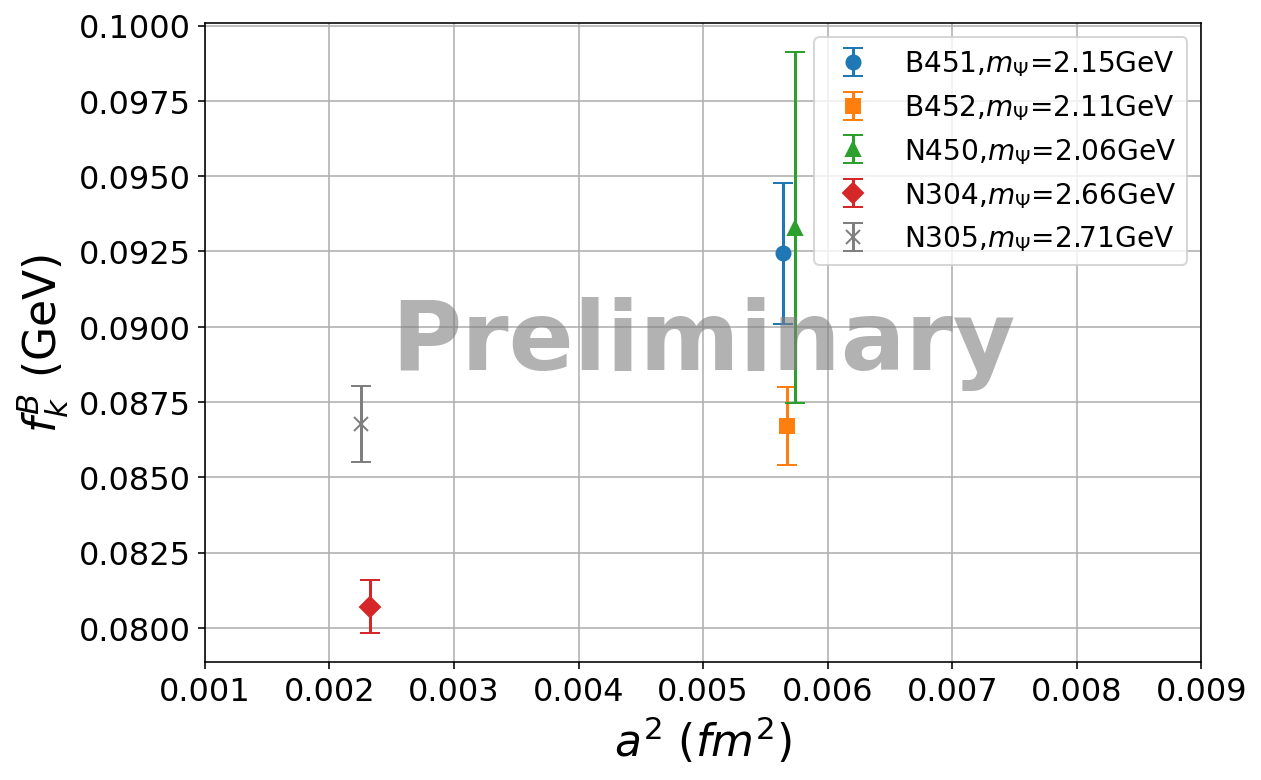}
  \end{minipage}\hfill
  \begin{minipage}[t]{0.49\textwidth}
    \centering
    \includegraphics[width=\linewidth]{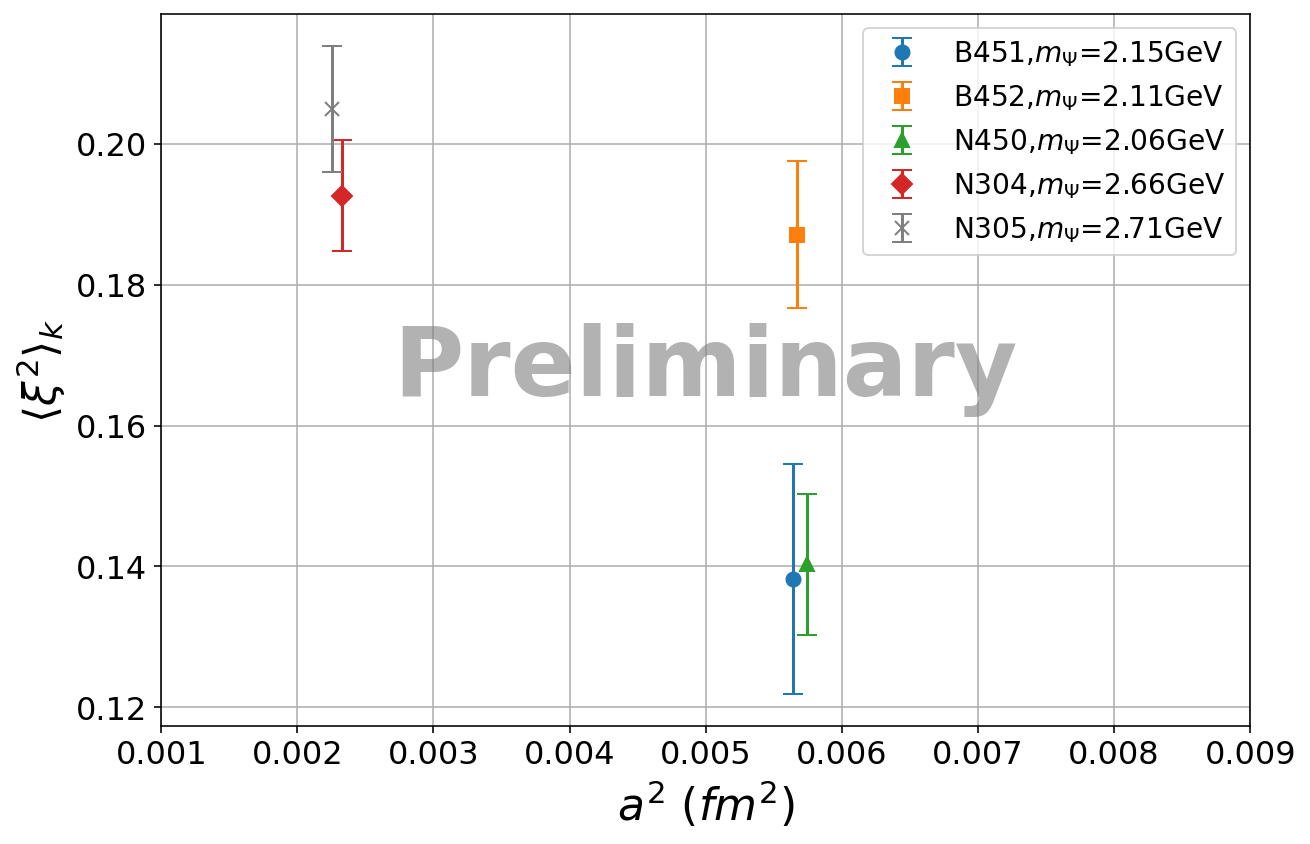}
  \end{minipage}
  \caption{Preliminary kaon results for the bare decay constant $f_K^{B}=f_K/Z_A^{2}$ (left) and the second Mellin moment (right) shown versus $a^2$ for different ensembles and heavy-quark masses.}
  \label{fig:decay_mom2_a2}
\end{figure}

Figure~\ref{fig:HadronicAmplitudes} shows representative one-loop HOPE fits for the kaon.
The fits reproduce the $\tau/a$ dependence of the ratios in the even and odd sectors, and therefore allow the determination of the first three nontrivial Mellin moments at the renormalization scale $\mu=2~\mathrm{GeV}$.
Figures~\ref{fig:mom13_a2} and~\ref{fig:decay_mom2_a2} summarize preliminary ensemble-by-ensemble results for the kaon moments and the kaon decay constant entering the normalization.
Final results will require extrapolations to the twist-2 and continuum limits, as well as to the physical pion and kaon masses.
A quenched continuum-limit result for the pion fourth Mellin moment, which serves as a benchmark for the HOPE framework, is discussed in Sec.~\ref{sec:pion4th}.

\section{First continuum limit of fourth Mellin moment}

\begin{figure}[htb]
\centering
\includegraphics[width=0.7\linewidth]{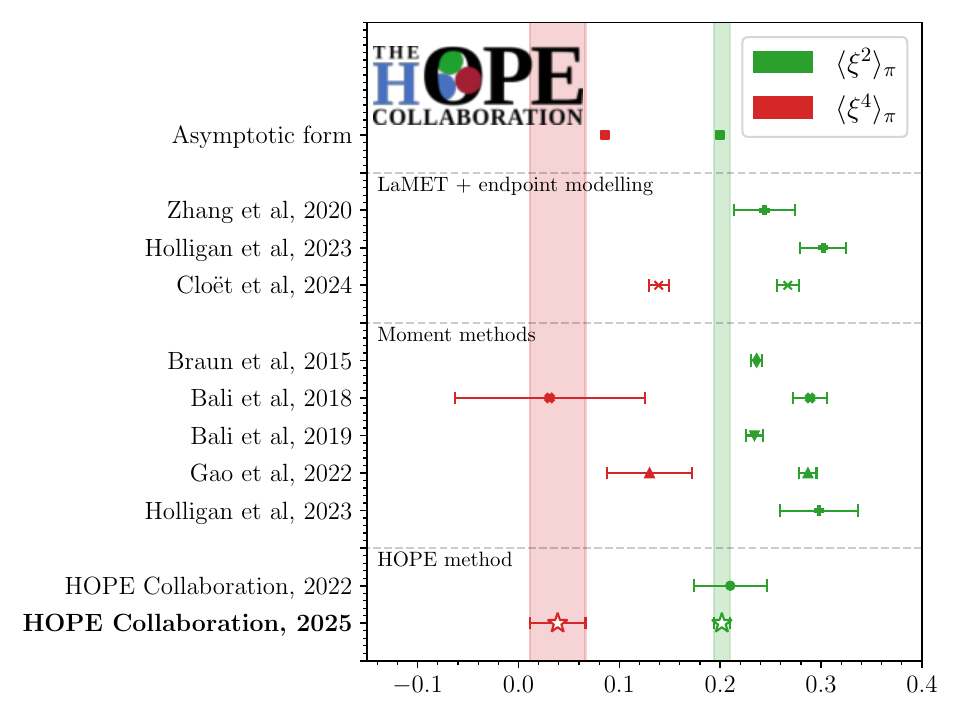}
\caption{Comparison of results obtained in Ref.~\cite{Detmold:2025lyb} to other determinations~\cite{Braun:2015axa,Bali:2018spj,RQCD:2019osh,Zhang:2020gaj,Detmold:2021qln,Gao:2022vyh,Cloet:2024vbv} of the second and the fourth  Mellin moments of the pion LCDA.}
\label{fig:quenched_comparison}
\end{figure}

An explicit realization of the required extrapolation to the twist-2, continuum limit can be found in Ref.~\cite{Detmold:2025lyb}, where the second and fourth Mellin moments of the pion are determined in the quenched approximation at quark masses corresponding to a pion mass of $m_\pi=550$~MeV using the HOPE method. Reference ~\cite{Detmold:2025lyb} constitutes the first continuum limit determination of the fourth Mellin moment. The results are $\langle\xi^2\rangle_\pi = 0.202(8)(9)$ and $\langle\xi^4\rangle_\pi = 0.039(28)(11)$, where the first error indicates the combined statistical and systematic uncertainty from the analysis and the second indicates the uncertainty from working with Wilson coefficients computed to next-to-leading order. These results are presented in $\overline{\mathrm{MS}}$  scheme at a renormalization scale of $\mu^2 = 2$ GeV. The result for the second Mellin moment are in good agreement with the previous determination of this quantity using the HOPE method~\cite{Detmold:2021qln}, and, after accounting for systematic effects arising from quark mass differences and the quenching approximation, also in good agreement with other existing calculations in the literature~\cite{Braun:2015axa,Bali:2018spj,RQCD:2019osh,Zhang:2020gaj,Detmold:2021qln,Gao:2022vyh,Cloet:2024vbv}. A summary of these calculations is shown in Fig.~\ref{fig:quenched_comparison}.

\label{sec:pion4th}
\section{Conclusions and outlook}

We reported the status of HOPE determinations of LCDA moments using an analysis procedure in the time-momentum representation. 
For the kaon, our dynamical ($N_f=2+1$) analysis shows sensitivity to the first three nontrivial Mellin moments, including a nonzero odd-channel signal, based on one-loop descriptions of the reconstructed ratios in selected kinematics.
For the pion, we discussed the first continuum-limit determination of the fourth Mellin moment in the quenched approximation.

Ongoing work adds statistics, additional heavy masses, and ensembles to enable controlled extrapolations to the twist-2 and continuum limits and to the physical pion and kaon masses.

\section*{Acknowledgments}
AVG is supported by the National Science Foundation under Grant No.~PHY-240227. The work of YZ is supported by the U.S. Department of Energy, Office of Science, Office of Nuclear Physics through Contract No.~DE-AC02-06CH11357. WD, MG and RJP are supported in part by the U.S. Department of Energy, Office of Science under Contract No.~DE-SC0011090 and by the SciDAC5 Award No.~DE-SC0023116. RJP has been supported in part by Simons Foundation Grant No. 994314 (Simons Collaboration on Confinement and QCD Strings). SPAC and CJDL are supported by the Taiwanese NSTC through grants 112-2112-M-A49-021-MY3 and 114-2123-M-A49-001-SVP.

We thank the CLS initiative for providing the gauge ensembles used in this study. Some of the numerical simulations were performed on the supercomputer Fugaku at the RIKEN Center for Computational Science(HPCI project hp220312 and hp230466), using the \texttt{Bridge++} \cite{Aoyama:2023tyf}. The authors thank ASRock Rack Inc. for their support of the construction of an Intel Knights Landing cluster at National Yang Ming Chiao Tung University, where the numerical calculations were performed. The authors thankfully acknowledge the computer resources at MareNostrum and the technical support provided by BSC (RES-FI-2023-1-0030). The authors gratefully acknowledge the support of ASGC (Academia Sinica Grid Computing Center, AS-CFII-114-A11, NSTC (NSTC 113-2740-M-001-007) for provision of computational resources. The research reported in this work made use of computing facilities of the USQCD Collaboration, which are funded by the Office of Science of the U.S. Department of Energy. Help from Balint Jo\'o in tuning Chroma is acknowledged.

\end{document}